\documentclass[sigconf]{acmart}

\setcopyright{none}
\renewcommand\footnotetextcopyrightpermission[1]{}
\AtBeginDocument{%
  }

\usepackage{microtype}
\usepackage{listings}
\usepackage{xspace}
\usepackage{pifont}
\usepackage{tikz}
\usetikzlibrary{shapes.geometric,arrows.meta,positioning,fit,backgrounds,calc}
\newcommand{\cmark}{\ding{51}}
\newcommand{\xmark}{\ding{55}}

\newcommand{\sys}{\textsc{Sandlock}\xspace}

\begin{document}

\title{Sandlock: Confining AI Agent Code with Unprivileged Linux Primitives}

\acmConference[Agentic OS Workshop]{Agentic OS Workshop, ASPLOS 2026}{March 23, 2026}{Pittsburgh, USA}

\author{Cong Wang}
\authornote{Corresponding author.}
\affiliation{%
	\institution{Multikernel Technologies, Inc.}
	\country{USA}
}
\email{cwang@multikernel.io}

\author{Yusheng Zheng}
\affiliation{%
	\institution{University of California, Santa Cruz}
	\country{USA}
}
\email{yzhen165@ucsc.edu}

\begin{abstract}
	AI agents increasingly run untrusted code on developer machines: shell commands generated by language models, third-party scripts retrieved at runtime, and tool plugins of unknown provenance. Existing isolation mechanisms impose tradeoffs that fit this workload poorly: containers and microVMs add privilege, image-management, and startup costs, while ad-hoc process controls and wrappers (e.g.\ \texttt{chroot}, \texttt{ulimit}) provide weak guarantees and little syscall-level control. \sys is a lightweight Linux process sandbox organized around a simple split: static, input-independent policy is compiled into kernel-enforced rules, while a narrow supervisor handles runtime-dependent decisions and virtualized effects. This split lets \sys enforce filesystem, network, IPC, and syscall policies without root, cgroups, images, or mandatory namespaces. It also supports dynamic network decisions, HTTP-level access control, TOCTOU-safe inspection of \texttt{execve} arguments, and reversible filesystem effects. On our workstation, \sys adds roughly 5\,ms of startup overhead and runs Redis at bare-metal throughput (within measurement noise); its pipeline operator further supports per-stage confinement for separating data, network, and untrusted-content capabilities. \sys is available at \url{https://github.com/multikernel/sandlock}.
\end{abstract}


\maketitle

\section{Introduction}

AI coding assistants and autonomous agents~\cite{swe-agent,openhands,claudecode} routinely execute code that no human has reviewed: shell commands proposed by the model, scripts downloaded from package registries, build steps invoked through MCP-style tool servers~\cite{mcp-tools}, and plug-ins from third-party marketplaces. Each invocation is a potential supply-chain exposure on the developer's workstation: an unintended \texttt{rm -rf}, access to an SSH key, a network call to an attacker-controlled host, or a \texttt{curl | sh} pipeline invoked by a model-generated build script.

In principle, the operating system has the building blocks to confine such code. In practice, off-the-shelf options make tradeoffs that fit this workload poorly. Containers~\cite{docker} and microVMs~\cite{firecracker} provide familiar isolation boundaries, but they bring image management, root or rootless configuration~\cite{namespaces,docker-rootless}, KVM requirements, and startup costs that are visible when an agent runs many short-lived commands. gVisor~\cite{gvisor} interposes a user-mode kernel, improving isolation at the cost of compatibility and syscall overhead for build tools. Ad-hoc controls such as \texttt{ulimit}, privileged \texttt{chroot}, and launcher tools such as \texttt{firejail}/\texttt{bubblewrap}~\cite{firejail,bubblewrap} reduce deployment cost but offer limited syscall control, weak filesystem semantics, and no host- or HTTP-level network policy.

This workload needs a tool tailored to the agent workstation: \emph{unprivileged}, \emph{low-latency}, \emph{filesystem-aware}, \emph{network-aware} at the level of hosts and HTTP routes, and \emph{programmable} so that a host application can enforce policies that depend on the program's runtime behavior (e.g.\ revoke network access once the workload transitions from setup to running untrusted code, a transition observable at \texttt{execve}).

The design problem is not the absence of Linux isolation primitives, but deciding which policies should be enforced where. Static invariants, such as readable path prefixes, writable workspaces, TCP ports, IPC scope, and unconditional syscall denials, should stay in kernel-enforced rules. Policies that depend on runtime values, such as the resolved destination of a \texttt{connect}, the \texttt{argv} of an \texttt{execve}, an HTTP method/path, or a copy-on-write filesystem effect, require controlled supervisor involvement. \sys makes this split explicit.

\sys is a Rust process sandbox that realizes this split using Landlock~\cite{landlock} for static filesystem, TCP-port, and IPC scope; seccomp-bpf~\cite{seccomp} for unconditional syscall filtering; and seccomp user notification~\cite{seccomp-unotify} with \texttt{pidfd\_getfd}~\cite{pidfd} for runtime-dependent decisions and on-behalf operations.

Rather than exposing these mechanisms independently, \sys presents a single policy surface: static rules are pushed into Landlock and seccomp-bpf whenever possible, and only runtime-dependent operations reach the supervisor. The same split supports HTTP-level network policy, reversible filesystem effects, and pipeline stages with different confinements. On our workstation, \sys adds $\sim$5\,ms of startup overhead ($\sim$6\,ms wall, 44$\times$ faster end-to-end than Docker on our setup) and no measurable Redis throughput overhead (within run-to-run noise).

This paper makes the following contributions:
\begin{itemize}
	\item A policy model and design for unprivileged agent sandboxing that separates static kernel-enforced policy from runtime-dependent supervisor decisions, while preserving a uniform configuration surface (Sections~\ref{sec:policy}--\ref{sec:design}).
	\item A TOCTOU-safe runtime policy mechanism (\texttt{policy\_fn}) that exposes syscall events, including \texttt{argv} for \texttt{execve}, to host code without violating kernel re-read semantics (Section~\ref{sec:policyfn}).
	\item Two applications of the split enforcement model: a seccomp-driven copy-on-write workspace that captures filesystem writes for commit/abort/dry-run \emph{without} mount namespaces or root, and a COW-fork primitive that sustains $\sim$1{,}900 sandboxed forks/s for map-reduce workloads (Section~\ref{sec:cow}).
	\item A preliminary evaluation showing $\sim$5\,ms startup overhead, bare-metal Redis throughput within measurement noise, and lower p99 latency than Docker (Section~\ref{sec:eval}).
\end{itemize}

\sys is open source at \url{https://github.com/multikernel/sandlock}.

\section{Motivation}
\label{sec:motivation}

\subsection{The Agent Workstation Threat Model}

These systems interleave LLM reasoning with shell commands: editing files, installing packages, running tests, and calling network APIs. We assume the sandboxed process may execute attacker-influenced code or consume attacker-controlled data, including package scripts, downloaded files, tool outputs, and generated commands. The host kernel, the \sys supervisor, and the embedding agent runtime are trusted; kernel vulnerabilities, side channels, privileged local attackers, and deliberate exhaustion of global kernel resources are out of scope. Five properties of this workload shape our threat model:

\paragraph{Untrusted code is short-lived and frequent.} A single agent task may invoke \texttt{npm install}, \texttt{pytest}, \texttt{curl}, or generated Python snippets dozens of times per minute. Per-invocation startup overhead in the hundreds of milliseconds (typical of containers) is visible in human-facing latency.

\paragraph{The host is a developer machine.} Sandboxes run alongside the developer's editor, SSH keys, and source trees. Confinement must be unprivileged (developer accounts rarely have free use of \texttt{sudo}) and must default-deny the rest of the home directory.

\paragraph{Network access is essential but dangerous.} Agents legitimately need to call LLM APIs, package registries, and documentation servers, but exfiltration to attacker-controlled hosts must be blocked. Allowlisting at the IP/port layer is necessary but not sufficient: a model that legitimately calls \texttt{api.openai.com:443} can also call attacker-controlled endpoints through the same allowed IP. Useful policies often live at the HTTP method, host, and path level.

\paragraph{Agent tasks compose stages of asymmetric trust.} A growing class of agent failures comes not from untrusted \emph{code} but from untrusted \emph{data}: prompt injection~\cite{camel}, where instructions embedded in a file, web page, or tool output hijack the agent's own LLM. The danger materializes when a single context combines access to private data, the ability to communicate externally, and exposure to untrusted content: the ``lethal trifecta''~\cite{lethaltrifecta}. Process-level confinement of one sandbox cannot break this combination, because a legitimate agent step often needs all three. This motivates splitting a task into differently confined stages: the stage exposed to untrusted content has no network, and the stage with network never sees it. Dual-LLM and capability-separation patterns~\cite{dualllm,camel} prescribe this decomposition but assume an enforcement substrate that unprivileged sandboxes do not provide.

\paragraph{Agent behavior is generated, not known in advance.} A static policy must be fixed before the agent decides what to run, yet the safe policy often depends on information that exists only at runtime. An \texttt{npm install} step legitimately needs network access and writes under \texttt{node\_modules} that the test step it sets up should \emph{not} retain; the transition between the two is observable (the \texttt{execve} of the test runner) but not expressible as a single static ruleset over one invocation. The decision can also hinge on a syscall argument (the \texttt{argv} of a model-generated command, the resolved destination of a \texttt{connect}) that no allowlist could enumerate ahead of time. Finally, the agent runtime embedding the sandbox usually knows more than the static config did: which tool is being invoked, what the previous step did, whether this sub-task was ever supposed to spawn a process at all. Confinement must therefore be \emph{programmable}: the host must be able to observe syscall events and tighten policy live, without re-launching the workload.

\subsection{Why Existing Mechanisms Fall Short}

Table~\ref{tab:comparison} compares \sys against representative isolation mechanisms.

\begin{table*}[t]
	\centering
	\caption{Comparison of unprivileged isolation mechanisms for the agent workstation.}
	\label{tab:comparison}
	\small
	\begin{tabular}{lccccc}
		\toprule
		Property             & \sys             & Docker (rootless) & Firecracker & gVisor       & firejail/bwrap \\
		\midrule
		No root required     & \cmark           & \cmark$^{\dagger}$ & \xmark      & \cmark       & \cmark         \\
		No image build       & \cmark           & \xmark           & \xmark      & \xmark       & \cmark         \\
		Startup latency      & $\sim$6\,ms      & $\sim$300\,ms    & $\sim$100\,ms & $\sim$200\,ms & $\sim$30\,ms   \\
		Per-syscall control  & seccomp-bpf      & seccomp default  & N/A         & user kernel  & seccomp        \\
		Filesystem ACL       & Landlock + COW   & overlay + bind   & block-level & overlay      & bind mounts    \\
		Host-level network   & on-behalf path   & netns            & TAP         & netns        & netns          \\
		HTTP-level network   & method/host/path$^{\ddagger}$ & N/A              & N/A         & N/A          & N/A            \\
		Runtime callback     & \texttt{policy\_fn} & N/A           & N/A         & N/A          & N/A            \\
		Pipeline composition & \texttt{|} operator & N/A            & N/A         & N/A          & N/A            \\
		\bottomrule
	\end{tabular}

	\smallskip
	{\scriptsize $^{\dagger}$Rootless Docker requires user-namespace support and \texttt{/etc/subuid} configuration~\cite{docker-rootless}. $^{\ddagger}$Method/host/path control over HTTPS requires installing a sandbox CA in the workload; without it, HTTPS is governed by pinned-endpoint allowlisting.}
\end{table*}

Containers and microVMs are designed for service deployment; their image and namespace machinery can impose deployment overhead disproportionate to a single 50-millisecond Python invocation. gVisor's user kernel adds compatibility and performance costs for build tools with subtle syscall dependencies. Tools such as \texttt{firejail} and \texttt{bubblewrap} stop short of programmable runtime policies and lack a host-level network ACL. None of these tools natively expose HTTP method/path-level access control, a useful granularity for agent network policy.

\subsection{Goals}

We extracted five design goals from the workload above:

\begin{itemize}
	\item \textbf{G1 Unprivileged.} No root, cgroups, or pre-provisioned namespaces.
	\item \textbf{G2 Low latency.} Sub-10\,ms startup; near-native runtime overhead.
	\item \textbf{G3 Programmable.} The host application can attach a callback that observes syscall events and adjusts policy live, without re-launching the workload.
	\item \textbf{G4 Composable.} Multi-stage workloads (planner/executor, mapper/reducer) should pipe data between stages whose confinements differ, so that the capability separation a prompt-injection-resistant decomposition relies on is enforced by the kernel rather than trusted to the LLM's judgment.
	\item \textbf{G5 Reversible.} Filesystem effects can be captured, previewed (\texttt{--dry-run}), and either committed or discarded on exit.
\end{itemize}

\section{Policy Model}
\label{sec:policy}

\sys exposes policy at the level of a command or pipeline stage. A policy describes which host resources the stage may observe, which effects it may produce, and which runtime events the embedding application may inspect before allowing the stage to continue. The model is intentionally smaller than a container configuration: it does not describe images, namespaces, or virtual networks, but the permissions and effects relevant to short-lived agent actions.

\subsection{Static Rules and Runtime Decisions}

The first distinction is between rules whose verdict is known before the child executes and rules whose verdict depends on syscall-time state. Static rules include readable and writable path prefixes, denied paths, TCP ports, IPC scope, and unconditional syscall denials. Runtime decisions include the resolved destination of a \texttt{connect}, the argument vector of an \texttt{execve}, an HTTP method/host/path, and filesystem effects that should be captured rather than applied directly. \sys requires runtime decisions to be made at a point where the relevant syscall can still be allowed, denied, or virtualized; if the decision cannot be made without a race or semantic mismatch, the policy fails closed.

\subsection{Policy Domains}

Filesystem policy grants read and write capabilities over host paths and can attach an effect action to writes: \texttt{COMMIT} merges captured changes, \texttt{ABORT} discards them, and \texttt{KEEP} leaves them for inspection. Network policy names endpoints by protocol and destination, with optional HTTP rules that further constrain method, host, and path. Resource policy bounds process count, memory, CPU share, and file descriptors. Runtime policy exposes selected syscall events to a host callback (\texttt{policy\_fn}), which may allow, deny, audit, or tighten the live policy for the remaining execution.

\subsection{Stage Composition}

For prompt-injection and data-flow isolation, policies compose across stages rather than only within one process tree. A pipeline is a sequence of stages connected by pipes; each stage has its own filesystem, network, resource, and runtime policy. This lets an agent author express capability separation directly: for example, one stage may read private data but lack network access, while another may call an external API but never receive that data except through an explicit pipe.

\section{\sys Design}
\label{sec:design}

\sys maps the policy model onto a split enforcement architecture. It follows one rule throughout: use the earliest enforcement point that can decide a policy soundly. If a decision is known before \texttt{execve}, \sys compiles it into Landlock or seccomp-bpf so the kernel enforces it without supervisor involvement. If a decision depends on syscall-time state, such as a resolved network address, an HTTP request, an \texttt{execve} argument vector, or a filesystem effect that should be captured rather than applied, \sys routes only that syscall class through seccomp user notification. If neither layer can enforce the policy without a race or a semantic mismatch, \sys denies the operation rather than silently weakening the policy.

\subsection{The Confinement Pipeline}

\begin{figure}[t]
	\centering
	\begin{tikzpicture}[
		node distance=0.18cm,
		cstep/.style={rectangle, draw, fill=blue!10, minimum width=4.6cm, minimum height=0.5cm, font=\small\ttfamily, anchor=west},
		pstep/.style={rectangle, draw, fill=orange!20, minimum width=4.6cm, minimum height=0.5cm, font=\small, anchor=west},
		]
		\node[pstep] (p1) at (0,0) {Parent: fork()};
		\node[cstep,below=of p1] (s1) {1. setpgid(0, 0)};
		\node[cstep,below=of s1] (s2) {2. chdir(cwd) [optional]};
		\node[cstep,below=of s2] (s3) {3. PR\_SET\_NO\_NEW\_PRIVS};
		\node[cstep,below=of s3] (s4) {4. Landlock: fs + net + IPC};
		\node[cstep,below=of s4] (s5) {5. seccomp filter (deny + notif)};
		\node[pstep,below=of s5] (p2) {Parent: recv notif fd, start supervisor};
		\node[cstep,below=of p2] (s6) {6. wait for ``ready'' from parent};
		\node[cstep,below=of s6] (s7) {7. close fds 3+};
		\node[cstep,below=of s7] (s8) {8. exec(cmd)};
		\end{tikzpicture}
		\caption{Confinement pipeline. Orange boxes execute in the parent, blue in the child. All steps are unprivileged.}
		\Description{A vertical flowchart showing parent fork, child confinement setup, supervisor startup, descriptor cleanup, and command execution.}
		\label{fig:pipeline}
	\end{figure}
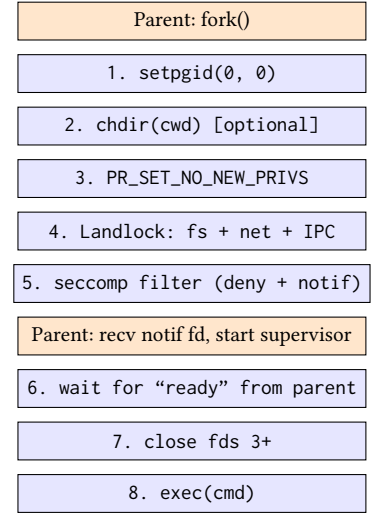

Confinement is applied in a fixed order after \texttt{fork()}, illustrated in Figure~\ref{fig:pipeline}. Each step is unprivileged and idempotent. The child blocks before \texttt{exec()} until the parent signals readiness, ensuring the supervisor is attached to the seccomp notification fd before any user code runs.

\subsection{Landlock as the Static Spine}

Landlock~\cite{landlock} provides the kernel-enforced ``static'' layer of policy. \sys uses it for constraints whose verdict is independent of syscall-time data: filesystem read/write scope, TCP port scope, and IPC boundaries. When a rule can be expressed this way, common operations are denied or allowed by the kernel without involving the supervisor. \sys requires Landlock ABI~6 or later (Linux~6.12+), which supplies filesystem rules (ABI~1), TCP-port rules (ABI~4), and abstract-socket and signal scoping for IPC (ABI~6)~\cite{landlock-man}; it queries the runtime ABI at startup and refuses to run below this floor rather than silently downgrading to a weaker policy. A \texttt{check} subcommand reports per-feature support (the evaluation kernel, 6.18, provides ABI~7).

\subsection{seccomp-bpf as the Filter}

After Landlock, \sys installs a seccomp-bpf filter that denies syscalls outside Landlock's scope, routes supervisor-visible syscalls to seccomp user notification (Table~\ref{tab:supervisor}), and lets the remaining calls proceed in the kernel.

The deny set is conservative by default. Because Landlock already covers filesystem, network, and IPC, the seccomp deny set focuses on syscalls that have no Landlock equivalent.

\subsection{The seccomp Notification Supervisor}

The dynamic layer of \sys is a user-notification supervisor, an async (tokio) task in the parent that dequeues notifications from the seccomp fd and replies with \texttt{Allow}, \texttt{Deny(errno)}, or \texttt{Continue} after performing an on-behalf operation. Table~\ref{tab:supervisor} summarizes the syscalls handled.

\begin{table}[t]
	\centering
	\caption{Syscalls intercepted by the seccomp notification supervisor.}
	\label{tab:supervisor}
	\small
	\begin{tabular}{ll}
		\toprule
		Syscall(s) & Supervisor responsibility \\
		\midrule
		\texttt{clone}/\texttt{fork}/\texttt{vfork}/\texttt{clone3} & Process count enforcement \\
		\texttt{mmap}/\texttt{munmap}/\texttt{brk} & Memory limit accounting \\
		\texttt{connect}, \texttt{send*} & IP allowlist + HTTP redirect \\
		\texttt{bind} & On-behalf bind \\
		\texttt{openat} & COW read redirect, write capture \\
		\texttt{unlinkat}/\texttt{mkdirat}/\texttt{renameat2} & COW write capture \\
		\texttt{exec*} & \texttt{policy\_fn} hold \\
		\texttt{getdents64} & COW directory merge \\
		\bottomrule
	\end{tabular}
\end{table}

For ``acting on behalf'' (e.g.\ to call \texttt{connect} from the child's address space using a parent-validated address), the supervisor uses \texttt{pidfd\_getfd}~\cite{pidfd} to dup the child's socket into the parent. This is TOCTOU-safe along two axes. The original syscall is never \texttt{Continue}d, so the kernel never re-reads the child's pointer arguments. Where an argument is itself passed by pointer (the destination \texttt{sockaddr} of \texttt{connect}/\texttt{sendto} or the payload of \texttt{sendmsg}), the supervisor reads it once, validates that copy, and issues the operation from its own copy, so a peer thread cannot substitute a different value after validation; each such read is bracketed by \texttt{SECCOMP\_IOCTL\_NOTIF\_ID\_VALID} so a read against a task that has exited or been replaced is rejected rather than trusted.

\subsection{Network Model}
\label{sec:net}

\sys uses a single endpoint allowlist that names \emph{protocol} $\times$ \emph{destination}. Rules may refer to concrete TCP, UDP, or ICMP endpoints, or to TCP ports when host identity is irrelevant. Concrete hostnames are resolved once at sandbox start and pinned for the child process, eliminating DNS-rebinding attacks. Wildcard hostnames are intentionally outside the model: every allowed domain must resolve to a concrete destination that the supervisor can check.

Two enforcement paths implement this model:

\paragraph{Direct path.} If TCP policy depends only on ports and no HTTP rule is active, Landlock enforces TCP-connect at the kernel level with no per-syscall overhead. UDP and ICMP, which Landlock does not cover, fall back to the on-behalf path when allowed.

\paragraph{On-behalf path.} Host-specific rules, HTTP rules, and non-TCP protocols route network syscalls through seccomp notification. The supervisor validates the destination against the resolved allowlist, then either performs the operation on behalf of the child or denies it.

\subsection{HTTP/HTTPS Access Control}

\sys refines endpoint policy with HTTP method, host, and path rules. For intercepted HTTP flows, the supervisor redirects the connection to a local proxy that parses request headers and applies the rule set before forwarding. HTTPS inspection is opt-in because it requires the workload to trust a sandbox CA; otherwise TLS flows are governed only by endpoint policy. HTTP rules also imply the corresponding endpoint permissions, so the policy does not require duplicate host-level configuration.

\subsection{Resource Limits Without Cgroups}

Without cgroups, traditional resource caps are unavailable. \sys reuses the supervisor for accounting:

\begin{itemize}
	\item \textbf{Memory.} The supervisor accounts for the virtual address space requested through \texttt{mmap}/\texttt{mremap}/\texttt{brk}/\texttt{shmget}; since resident pages cannot exceed mapped address space, this upper-bounds the footprint of those mappings. On overshoot the process is terminated.
	\item \textbf{Process count.} Fork-like syscalls are gated by a counter; thread creation (\texttt{CLONE\_THREAD}) is excluded, and excess processes receive \texttt{EAGAIN}.
	\item \textbf{CPU throttling.} An optional task issues \texttt{SIGSTOP}/\texttt{SIGCONT} to the child's process group at a duty cycle proportional to \texttt{max\_cpu}.
	\item \textbf{File descriptors} use a standard \texttt{RLIMIT\_NOFILE}.
\end{itemize}

The trade-off is that these limits are enforced \emph{cooperatively} via syscall interception rather than at allocation time inside the kernel, and the memory bound is on \emph{address space} requested rather than resident pages: growth that bypasses these syscalls (notably main-thread stack growth, which \texttt{RLIMIT\_STACK} governs instead) is not counted, and for a pure address-space cap \texttt{RLIMIT\_AS} is a simpler kernel-enforced equivalent. We route memory through the supervisor so the same accounting can also virtualize \texttt{/proc/meminfo}. For accidental over-allocation by trusted-but-buggy code these limits suffice; for adversarial workloads aiming to exhaust kernel memory through corner-case syscalls, cgroups remain stronger~\cite{agentcgroup}.

\section{Implementation}
\label{sec:impl}

\sys is implemented in $\sim$7K lines of Rust. Its core library constructs Landlock rules, compiles seccomp filters, runs the async supervisor, and implements the COW, pipeline, and runtime-callback mechanisms. A small command-line frontend and FFI layer expose the same policy model to shell and Python callers.

\noindent The remainder of this section describes the implementation details most relevant to safety and portability: the runtime callback boundary, the unprivileged COW backends, and the fork/pipeline runtimes.

\subsection{Programmable Policies via \texttt{policy\_fn}}
\label{sec:policyfn}

\begin{figure}[t]
\begin{lstlisting}
def on_event(event, ctx):
  if event.syscall == "execve":
    # argv is an observation signal, not a boundary
    if event.argv_contains("curl"):
      return "audit"                  # flag; name matching is evadable
    ctx.restrict_network([])          # real control: revoke network
    ctx.deny_path("/etc/shadow")      #   and tighten fs scope
  if event.category == "file":
    return "audit"                    # allow + flag
  return 0                            # allow
	\end{lstlisting}
		\caption{Programmable policy callback in Python.}
		\Description{Python pseudocode for a policy callback that audits curl execution, tightens network and filesystem policy, and audits file events.}
		\label{fig:policyfn}
	\end{figure}

The runtime policy hook from Section~\ref{sec:policy} is implemented as a callback, \texttt{policy\_fn}, registered in Rust or Python (Figure~\ref{fig:policyfn}). The callback receives an \texttt{Event} containing the syscall name, category, PID/PPID, selected network destination fields, and \texttt{argv} for \texttt{execve}. It also receives a \texttt{Context} for live policy updates.

Argv inspection is an observation and policy-tightening signal, not a containment boundary: matching a binary name cannot stop adversarial code, which can rename or reimplement the tool. Containment comes from Landlock, seccomp, and the network ACL; \texttt{policy\_fn} adds the ability to detect a phase transition (e.g.\ setup~$\rightarrow$~untrusted execution) and tighten those kernel-enforced policies live for the rest of the run.

\paragraph{TOCTOU safety.} Per \texttt{seccomp\_unotify(2)}, after the supervisor returns \texttt{Continue}, the kernel re-reads user-memory pointers. A naive implementation that copies \texttt{argv} into the event and then \texttt{Continues} the syscall would race with a peer thread or process aliasing the same memory. \sys avoids this race by keeping path strings out of runtime events: path-based control remains in static Landlock rules, which are TOCTOU-immune, and live path restrictions are represented as policy updates rather than inspected syscall arguments. For \texttt{argv}, which the host may need to inspect, \sys freezes every task that could alias the child's address space before reading arguments or continuing an \texttt{execve}. It enumerates sibling threads and peer processes from \texttt{/proc} and its process index, then stops them with ptrace seize/interrupt requests. Fork-like syscalls (\texttt{clone}, \texttt{fork}, \texttt{vfork}, and \texttt{clone3}) are traced for one ptrace creation event so children are registered before user code runs; for \texttt{clone3}, \sys reads flags from the clone-argument structure. If freezing or creation tracking is unavailable (e.g.\ Yama blocks ptrace), the syscall is denied with \texttt{EPERM} rather than silently relaxed.

The \emph{held syscalls}, those that block until the callback returns, are \texttt{execve}, \texttt{connect}, \texttt{sendto}, \texttt{sendmsg}, \texttt{sendmmsg}, \texttt{bind}, and \texttt{openat}. Verdicts are: \texttt{0}/\texttt{False} = allow, \texttt{True}/\texttt{-1} = deny (\texttt{EPERM}), positive integer = deny with that errno, \texttt{"audit"} = allow + flag.

\subsection{Copy-on-Write Workspace Backends}
\label{sec:cow}

\sys provides two COW backends, both unprivileged:

\paragraph{Seccomp COW (default).} The supervisor intercepts filesystem syscalls and redirects writes into an upper directory while resolving reads against upper and lower layers. Directory reads are merged in the supervisor. No mount namespace is needed.

\paragraph{BranchFS.} BranchFS,\footnote{\url{https://github.com/multikernel/branchfs/}} a dedicated copy-on-write filesystem, captures writes at the filesystem layer rather than through per-syscall interception. \sys drives the branch lifecycle with \texttt{ioctl}s on a control file (create on entry, commit or abort on exit), and an optional storage quota bounds the branch, surfaced to the child as \texttt{ENOSPC}. Like the seccomp backend it needs no mount namespace in the child, but it keeps the write path inside the filesystem, avoiding a supervisor round-trip per filesystem syscall.

Both backends expose the same effect interface: captured filesystem updates can be committed, discarded, or kept after the sandbox exits.

\subsection{COW Fork Runtime}

The COW-fork primitive runs initialization once and then forks $N$ confined workers so they share initialized pages copy-on-write. The runtime captures per-worker output and can pipe merged results to a separately confined reducer. On our setup this sustains $\sim$1{,}900 forks/s, enabling map-reduce patterns directly inside an agent process.

\subsection{Pipeline Runtime}

\begin{figure}[t]
\begin{lstlisting}
trusted = Sandbox(
    fs_readable=["/usr", "/lib", "/opt/data"])
restricted = Sandbox(
    fs_readable=["/usr", "/lib"])  # no /opt/data
result = (
    trusted.cmd(["cat", "/opt/data/secret.csv"])
  | restricted.cmd(["tr", "a-z", "A-Z"])
).run()
	\end{lstlisting}
		\caption{Pipeline of heterogeneous sandboxes.}
		\Description{Python pseudocode composing a trusted stage with data access and a restricted stage connected by a pipe.}
		\label{fig:stage-pipeline}
	\end{figure}

The stage-composition policy is implemented by connecting sandboxed stages with kernel pipes. Each stage carries its own confinement, allowing the author to separate private-data access, network access, and exposure to untrusted content across different process trees. Figure~\ref{fig:stage-pipeline} shows the pattern.

Because each stage is its own \sys process tree, a compromise of \texttt{tr} cannot read \texttt{/opt/data}: only the upstream \texttt{cat} could, and its only output channel is the pipe. The same structure enforces the capability boundary that dual-LLM and CaMeL-style decompositions~\cite{dualllm,camel} assume: in the planner/executor split, the stage exposed to untrusted content can be denied network and the stage with network can be denied data access, so a hijacked stage holds no capability it was not granted. Two caveats bound this claim. First, \sys bounds stage \emph{capabilities}, not pipe \emph{content}: a compromised upstream stage can still shape bytes consumed downstream, so data-provenance defenses~\cite{camel} remain complementary. Second, sound decompositions remain the author's responsibility: \sys kernel-enforces the separation specified, but does not prove that the untrusted-content stage never needs network or that the networked stage never needs private data.

\section{Preliminary Evaluation}
\label{sec:eval}

\begin{figure*}[t]
	\centering
	\includegraphics[width=\textwidth]{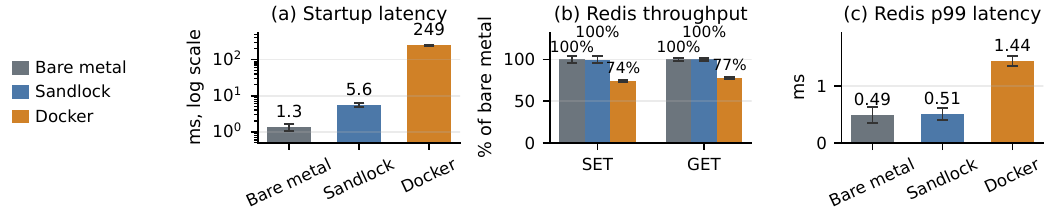}
	\caption{Preliminary benchmark results. \sys preserves low startup latency and near-bare-metal Redis throughput while avoiding Docker's per-invocation and tail-latency costs on this workstation.}
	\Description{Benchmark plots comparing bare metal, Sandlock, and Docker for echo startup latency, Redis throughput, Redis tail latency, and fork rate.}
	\label{fig:benchmarks}
\end{figure*}

\paragraph{Setup.} Experiments run on a typical Linux workstation: an AMD Ryzen 5 5500U (6~cores / 12~threads, up to 4.0\,GHz) with 8\,GB of RAM and NVMe SSD storage, running Pop!\_OS 24.04 LTS (an Ubuntu 24.04 derivative) on Linux 6.18 with unprivileged user namespaces enabled. Startup latency is the median of 20 runs (after a warm-up); Redis throughput and tail latency are the median of 5 runs of \texttt{redis-benchmark} (100{,}000 requests, 50 concurrent clients, 256\,B values), and error bars in Figure~\ref{fig:benchmarks} show $\pm$1 standard deviation. We report bare-metal, \sys (default static rules), and Docker for context. We benchmark rootful Docker here, the common deployment and the more favorable baseline for Docker; Table~\ref{tab:comparison} instead compares \emph{features} against rootless Docker, the unprivileged competitor relevant to \sys's no-root claim.

\paragraph{Effectiveness.} Before measuring overhead, we check that the rules and supervisor actually confine the workload. Table~\ref{tab:effectiveness} reports deterministic deny/allow cases drawn from the threat model (Section~\ref{sec:motivation}); each is run against \sys and the observed outcome recorded. Reads and writes outside the granted scope are denied while in-scope access succeeds; a \texttt{connect} to a host outside the endpoint allowlist is refused while an allowlisted endpoint connects; and a stage not granted a private directory cannot read it even though a sibling stage can, the per-stage capability separation the pipeline relies on. Cooperative resource caps also hold: process creation past the limit returns \texttt{EAGAIN}, and an allocation past the memory cap is blocked (the process is terminated rather than over-committing). All ten checks behaved as the policy requires.

\begin{table}[t]
	\centering
	\caption{Effectiveness micro-evaluation: deterministic deny/allow checks mapped to the threat model. Each row is the outcome observed when the case is run against \sys.}
	\label{tab:effectiveness}
	\small
	\begin{tabular}{lll}
		\toprule
		Attempted action & Policy & Outcome \\
		\midrule
		Read outside read scope        & default-deny      & denied \\
		Read inside read scope         & granted           & allowed \\
		Write outside write scope      & default-deny      & denied \\
		Write inside write scope       & granted           & allowed \\
		Connect to non-allowlisted host& allowlist         & refused \\
		Connect to allowlisted host    & allowlist         & allowed \\
		Untrusted stage reads private data & no grant      & denied \\
		Data stage reads private data  & granted           & allowed \\
		Fork past process cap          & \texttt{-P 4}     & \texttt{EAGAIN} \\
		Alloc past memory cap          & \texttt{-m 64M}   & blocked \\
		\bottomrule
	\end{tabular}
\end{table}

\paragraph{Startup overhead.} Figure~\ref{fig:benchmarks}(a) shows \texttt{/bin/echo} wall time. \sys adds $\sim$5\,ms over bare metal, dominated by Landlock ruleset construction and seccomp-fd setup; Docker is roughly 44$\times$ slower due to image and namespace setup.

\paragraph{Throughput.} We benchmark Redis 7 with \texttt{redis-benchmark -n 100000} (Figure~\ref{fig:benchmarks}(b,c)). \sys matches bare-metal throughput on both SET and GET to within run-to-run variance (medians over 5 runs of 75.2k vs.\ 75.5k\,rps for SET and 74.2k vs.\ 73.9k\,rps for GET), with p99 latency close to bare metal (0.51 vs.\ 0.49\,ms); Docker reaches only $\sim$76\% of bare-metal throughput and has roughly 3$\times$ the p99 tail latency. Any residual cost from the seccomp filter and the on-behalf bind for the listening socket falls below this workload's measurement noise.

\paragraph{COW fork.} A 1{,}000-fork run takes $\sim$530\,ms. This is $\sim$530\,$\mu$s/fork, or $\sim$1{,}900 forks/s. The cost is dominated by the \texttt{ptrace} creation handshake for \texttt{policy\_fn}-safe child registration; with \texttt{policy\_fn} disabled, fork-rate increases by another $\sim$30\%.

\paragraph{Network on-behalf.} The Landlock direct path (port-only policy) leaves traffic in the kernel and adds no per-syscall overhead. The on-behalf path is different: each send is mediated by the supervisor, adding a fixed per-message cost. For a 256\,B request/response echo with both endpoints sandboxed, the median round-trip rises from 20\,$\mu$s on bare metal to 55\,$\mu$s ($\sim$35\,$\mu$s added; p99 33\,$\mu$s vs.\ 114\,$\mu$s), whereas Docker, whose established connections stay in the kernel, matches bare metal (19\,$\mu$s median). This fixed cost is immaterial for the network workloads agents actually run, such as LLM API calls measured in 100\,ms--10\,s.

\paragraph{Compatibility.} Common build tools run under \sys once their scratch paths are granted: we ran \texttt{python3}, \texttt{make}, \texttt{node}, and \texttt{pytest} unmodified; Python native-extension imports also worked. The main tuning is to grant each tool its temporary directories (e.g.\ \texttt{/tmp} for compiler intermediates, \texttt{/dev/null}). A systematic compatibility study across the long tail of agent tooling is future work.

\section{Related Work}

\paragraph{Unprivileged Linux primitives.} \sys is built on top of Landlock~\cite{landlock}, seccomp-bpf~\cite{seccomp}, and seccomp user notification~\cite{seccomp-unotify}, with \texttt{pidfd\_getfd}~\cite{pidfd} for safe fd transfer. Capsicum~\cite{capsicum} is the closest spiritual ancestor: a capability-based unprivileged sandbox, originally for FreeBSD, later partially ported. Container runtimes embody the same split at a coarser grain: runc and Docker can forward selected syscalls to a userspace agent through seccomp notification~\cite{seccomp-unotify,oci-runtime-spec}. \sys's contribution is therefore not the use of notification (which is its kernel-documented purpose) but how the primitives are organized: a single policy surface spanning filesystem, network, IPC, and resources; a TOCTOU-safe runtime hook that exposes \texttt{execve} \texttt{argv} to host code; and unprivileged copy-on-write plus pipeline composition built on the same supervisor. Among unprivileged process sandboxes, \texttt{nsjail}~\cite{nsjail} and \texttt{minijail}~\cite{minijail} compose namespaces with seccomp-bpf, and Rust libraries such as birdcage~\cite{birdcage} provide embeddable static sandboxing APIs; these enforce static policy but expose neither a runtime callback nor host/HTTP-level network rules. BPF-LSM sandboxes such as bpfbox~\cite{bpfbox} provide programmable policy through eBPF, but loading LSM BPF generally requires \texttt{CAP\_BPF} or root~\cite{capabilities}, which \sys avoids. The classic hazards of syscall interposition (argument races and indirect paths) were catalogued by Garfinkel~\cite{garfinkel-traps}; \sys's TOCTOU handling (Section~\ref{sec:policyfn}) is designed against exactly these.

\paragraph{Container and microVM sandboxes.} Docker~\cite{docker} and rootless containers compose namespaces~\cite{namespaces}, cgroups~\cite{cgroups}, and seccomp profiles into reusable images; gVisor~\cite{gvisor} adds a user-mode kernel; Firecracker~\cite{firecracker} and Catalyzer~\cite{catalyzer} reduce microVM startup. These target multi-tenant service hosting, where image immutability and strong kernel isolation outweigh the startup cost. \sys targets the orthogonal workstation use case where startup cost matters and root is unavailable.

\paragraph{Process-internal isolation.} lwCs~\cite{lwc} introduce in-process protection domains for fast context switching; Dune~\cite{dune} uses VT-x for user-level kernels. These provide stronger isolation \emph{within} a process; \sys is concerned with confining \emph{a different} process tree.

\paragraph{Agent systems.} MVVM~\cite{mvvm} deploys agents in WebAssembly enclaves; Kgent~\cite{kgent} explores LLM-driven kernel extensions; AgentCgroup~\cite{agentcgroup} uses cgroups to characterize and control the OS resources AI agents consume; Agentry~\cite{agentry} mediates agent-to-external communication through a gateway. \sys is complementary: it confines the local execution side of an agent stack while gateways like Agentry mediate the communication side.

\paragraph{Prompt injection.} The dual-LLM pattern~\cite{dualllm} and CaMeL~\cite{camel} defend against prompt injection by structurally separating the component exposed to untrusted content from the component holding sensitive capabilities. These works define the decomposition; \sys provides an unprivileged, kernel-enforced substrate on which the separated stages run, via its pipeline operator and heterogeneous per-stage confinements.

\paragraph{COW and speculation.} OverlayFS~\cite{overlayfs} and UnionFS~\cite{unionfs} provide kernel-level COW; \sys's seccomp COW is a userspace alternative that requires no mount namespace. MBOX~\cite{mbox} layered a sandbox filesystem over a process via \texttt{ptrace} interposition with a commit step, much as \sys's seccomp COW captures writes for commit or abort; \sys uses seccomp notification rather than \texttt{ptrace} and shares one supervisor across its filesystem, network, and runtime-policy paths. Speculator~\cite{speculator} and TxOS~\cite{txos} explored OS-level transactions and output gating; \sys's \texttt{--dry-run} mode is a much narrower point in this design space, restricted to filesystem effects. Recent agent-oriented work revisits these primitives: Fork, Explore, Commit~\cite{forkexplorecommit} proposes speculative agent exploration with commit/rollback, matching \sys's COW-fork and workspace effects; ACRFence~\cite{acrfence} shows that checkpoint-based rollback can itself be attacked.

\section{Discussion and Future Work}

\paragraph{Threat-model boundaries.} \sys is unprivileged by design and therefore inherits the limits of unprivileged Linux. Kernel-level vulnerabilities, side channels, and resource-exhaustion attacks against shared kernel structures are out of scope. For multi-tenant adversarial workloads, a microVM remains the right answer.

\paragraph{Integration.} \sys supports two retrofit modes. In \emph{wrap-the-child}, the host forks, confines, and execs each command an agent already spawns, a drop-in for the shell and tool calls an agent issues, requiring no change to the agent loop. In \emph{self-confinement}, the static Landlock layer is applied to the calling process in place (a \texttt{confine()} call), so an agent runtime can drop its own filesystem and network privileges after startup; the dynamic supervisor, which needs a separate parent, remains a wrap-the-child facility. MCP tool servers are a natural fit for the first mode: a long-running server confines each tool invocation as a child under a per-tool policy, isolating untrusted tool execution without re-architecting the agent.

\paragraph{Beyond the local host.} HTTP-level ACLs are a useful step, but agentic workloads increasingly span tool servers (MCP) and remote APIs. We are exploring an integration in which \sys cooperates with an Agentry-style~\cite{agentry} gateway: the local sandbox enforces method/host/path; the gateway enforces semantic and per-tenant policies on the same flow.

\paragraph{Effect rollback.} \sys captures filesystem effects but not network effects. A natural extension is to buffer outbound HTTP requests through the proxy and release them only after the host commits the sandboxed stage, mirroring the speculator-style output gating literature~\cite{speculator} for the agent setting.

\paragraph{Composition with branching filesystems.} \sys's BranchFS backend already gives a single sandbox a copy-on-write branch, but its pipeline and COW-fork primitives do not yet exploit branching. A natural next step is to run each pipeline stage and each COW-fork worker on its own branch and merge the results on commit, giving multi-branch agents end-to-end isolation across both filesystem and process state.

\section{Conclusion}

We presented \sys, an unprivileged process sandbox tailored to the AI-agent workstation. Its central design is a split enforcement model: static policy is compiled into Landlock and seccomp-bpf, while a narrow seccomp-notification supervisor handles runtime-dependent decisions and virtualized effects. This structure gives \sys a programmable runtime policy surface (\texttt{policy\_fn}), HTTP-level ACLs, copy-on-write workspaces, COW-fork map/reduce, and pipeline composition while preserving $\sim$5\,ms startup overhead and near-native Redis performance. \sys is open source at \url{https://github.com/multikernel/sandlock}.

\bibliographystyle{ACM-Reference-Format}
\bibliography{references}

@misc{landlock,
  author    = {Salaün, Mickaël},
  title     = {Landlock {LSM}: Toward Unprivileged Sandboxing},
  howpublished = {Linux Security Summit Europe},
  year      = {2017},
  url       = {https://landlock.io/talks/2017-09-14_landlock-lss.pdf},
}

@misc{landlock-man,
  author = {{Linux man-pages Project}},
  title = {\texttt{landlock(7)} --- {Linux} Manual Page},
  howpublished = {\url{https://man7.org/linux/man-pages/man7/landlock.7.html}},
  year = {2025},
}

@misc{seccomp,
  author = {Drewry, Will and Edge, Jake},
  title = {\texttt{seccomp(2)} --- {Linux} Manual Page},
  howpublished = {\url{https://man7.org/linux/man-pages/man2/seccomp.2.html}},
  year = {2024},
}

@misc{seccomp-unotify,
  author = {Kerrisk, Michael},
  title = {\texttt{seccomp\_unotify(2)} --- {Linux} Manual Page},
  howpublished = {\url{https://man7.org/linux/man-pages/man2/seccomp_unotify.2.html}},
  year = {2024},
}

@misc{pidfd,
  author = {Brauner, Christian},
  title = {\texttt{pidfd\_getfd(2)} --- {Linux} Manual Page},
  howpublished = {\url{https://man7.org/linux/man-pages/man2/pidfd_getfd.2.html}},
  year = {2024},
}

@misc{capabilities,
  author = {{Linux man-pages Project}},
  title = {\texttt{capabilities(7)} --- {Linux} Manual Page},
  howpublished = {\url{https://man7.org/linux/man-pages/man7/capabilities.7.html}},
  year = {2025},
}

@article{unionfs,
  author = {Wright, Charles P. and Dave, Jay and Gupta, Puja and Krishnan, Harikesavan and Quigley, David P. and Zadok, Erez and Zubair, Mohammad Nayyer},
  title = {Versatility and {Unix} Semantics in Namespace Unification},
  journal = {ACM Transactions on Storage},
  volume = {2},
  number = {1},
  pages = {74--105},
  year = {2006},
}

@misc{overlayfs,
  author = {Brown, Neil},
  title = {Overlay Filesystem},
  howpublished = {\url{https://www.kernel.org/doc/Documentation/filesystems/overlayfs.txt}},
  year = {2016},
}

@inproceedings{speculator,
  author = {Nightingale, Edmund B. and Chen, Peter M. and Flinn, Jason},
  title = {Speculative Execution in a Distributed File System},
  booktitle = {Proceedings of the Twentieth ACM Symposium on Operating Systems Principles},
  series = {SOSP '05},
  year = {2005},
  pages = {191--205},
  publisher = {ACM},
  address = {Brighton, United Kingdom},
}

@inproceedings{txos,
  author = {Porter, Donald E. and Hofmann, Owen S. and Rossbach, Christopher J. and Benn, Alexander and Witchel, Emmett},
  title = {Operating System Transactions},
  booktitle = {Proceedings of the ACM SIGOPS 22nd Symposium on Operating Systems Principles},
  series = {SOSP '09},
  year = {2009},
  pages = {161--176},
  publisher = {ACM},
  address = {Big Sky, MT, USA},
}

@inproceedings{firecracker,
  author = {Agache, Alexandru and Brooker, Marc and Iordache, Alexandra and Liguori, Anthony and Neugebauer, Rolf and Piwonka, Phil and Popa, Diana-Maria},
  title = {{Firecracker}: Lightweight Virtualization for Serverless Applications},
  booktitle = {Proceedings of the 17th USENIX Symposium on Networked Systems Design and Implementation},
  series = {NSDI '20},
  year = {2020},
  pages = {419--434},
  publisher = {USENIX Association},
  address = {Santa Clara, CA, USA},
}

@inproceedings{catalyzer,
  author = {Du, Dong and Yu, Tianyi and Xia, Yubin and Zang, Binyu and Yan, Guanglu and Qin, Chenggang and Wu, Qixuan and Chen, Haibo},
  title = {{Catalyzer}: Sub-millisecond Startup for Serverless Computing with Initialization-less Booting},
  booktitle = {Proceedings of the 25th ACM International Conference on Architectural Support for Programming Languages and Operating Systems},
  series = {ASPLOS '20},
  year = {2020},
  pages = {467--481},
  publisher = {ACM},
  address = {Lausanne, Switzerland},
}

@misc{claudecode,
  author = {{Anthropic}},
  title = {{Claude Code}: An Agentic Coding Tool},
  howpublished = {\url{https://code.claude.com/docs/en/overview}},
  year = {2025},
}

@misc{mcp-tools,
  author = {{Model Context Protocol}},
  title = {Tools},
  howpublished = {\url{https://modelcontextprotocol.io/specification/draft/server/tools}},
  year = {2026},
  note = {Accessed: 2026-05-23},
}

@misc{namespaces,
  author = {Kerrisk, Michael},
  title = {Namespaces in Operation},
  howpublished = {LWN.net},
  url = {https://lwn.net/Articles/531114/},
  year = {2013},
}

@misc{cgroups,
  author = {Heo, Tejun},
  title = {Control Groups v2},
  howpublished = {\url{https://www.kernel.org/doc/Documentation/cgroup-v2.txt}},
  year = {2015},
}

@misc{docker,
  author = {Merkel, Dirk},
  title = {{Docker}: Lightweight {Linux} Containers for Consistent Development and Deployment},
  howpublished = {Linux Journal},
  year = {2014},
  url = {https://www.linuxjournal.com/content/docker-lightweight-linux-containers-consistent-development-and-deployment},
}

@misc{docker-rootless,
  author = {{Docker Inc.}},
  title = {Rootless Mode},
  howpublished = {\url{https://docs.docker.com/engine/security/rootless/}},
  year = {2026},
  note = {Accessed: 2026-05-23},
}

@misc{gvisor,
  author = {{Google}},
  title = {{gVisor}: Application Kernel for Containers},
  howpublished = {\url{https://gvisor.dev/}},
  year = {2024},
}

@misc{mvvm,
  author = {Yang, Yiwei and Hu, Aibo and Zheng, Yusheng and Zhao, Brian and Zhang, Xinqi and Xiang, Dawei and Chu, Kexin and Zhang, Wei and Quinn, Andi},
  title = {{MVVM}: Deploy Your {AI} Agents---Securely, Efficiently, Everywhere},
  howpublished = {arXiv preprint arXiv:2410.15894},
  year = {2024},
  doi = {10.48550/arXiv.2410.15894},
}

@inproceedings{lwc,
  author = {Litton, James and Vahldiek-Oberwagner, Anjo and Elnikety, Eslam and Garg, Deepak and Bhattacharjee, Bobby and Druschel, Peter},
  title = {Light-Weight Contexts: An {OS} Abstraction for Safety and Performance},
  booktitle = {Proceedings of the 12th USENIX Symposium on Operating Systems Design and Implementation},
  series = {OSDI '16},
  year = {2016},
  pages = {49--64},
  publisher = {USENIX Association},
  address = {Savannah, GA, USA},
}

@inproceedings{dune,
  author = {Belay, Adam and Bittau, Andrea and Mashtizadeh, Ali and Terei, David and Mazi\`{e}res, David and Kozyrakis, Christos},
  title = {Dune: Safe User-level Access to Privileged {CPU} Features},
  booktitle = {Proceedings of the 10th USENIX Symposium on Operating Systems Design and Implementation},
  series = {OSDI '12},
  year = {2012},
  pages = {335--348},
  publisher = {USENIX Association},
  address = {Hollywood, CA, USA},
}

@inproceedings{capsicum,
  author = {Watson, Robert N. M. and Anderson, Jonathan and Laurie, Ben and Kennaway, Kris},
  title = {Capsicum: Practical Capabilities for {UNIX}},
  booktitle = {Proceedings of the 19th USENIX Security Symposium},
  series = {USENIX Security '10},
  year = {2010},
  pages = {29--46},
  publisher = {USENIX Association},
  address = {Washington, DC, USA},
}

@inproceedings{swe-agent,
  author = {Yang, John and Jimenez, Carlos E. and Wettig, Alexander and Lieret, Kilian and Yao, Shunyu and Narasimhan, Karthik and Press, Ofir},
  title = {{SWE-agent}: Agent-Computer Interfaces Enable Automated Software Engineering},
  booktitle = {Advances in Neural Information Processing Systems 37 (NeurIPS)},
  year = {2024},
  publisher = {Curran Associates, Inc.},
  address = {Vancouver, BC, Canada},
  pages = {50528--50652},
  doi = {10.52202/079017-1601},
  url = {https://proceedings.neurips.cc/paper_files/paper/2024/hash/5a7c947568c1b1328ccc5230172e1e7c-Abstract-Conference.html},
}

@inproceedings{openhands,
  author = {Wang, Xingyao and Li, Boxuan and Song, Yufan and Xu, Frank F. and Tang, Xiangru and Zhuge, Mingchen and Pan, Jiayi and Song, Yueqi and Li, Bowen and Singh, Jaskirat and Tran, Hoang H. and Li, Fuqiang and Ma, Ren and Zheng, Mingzhang and Qian, Bill and Shao, Yanjun and Muennighoff, Niklas and Zhang, Yizhe and Hui, Binyuan and Lin, Junyang and Brennan, Robert and Peng, Hao and Ji, Heng and Neubig, Graham},
  title = {{OpenHands}: An Open Platform for {AI} Software Developers as Generalist Agents},
  booktitle = {The Thirteenth International Conference on Learning Representations},
  year = {2025},
  publisher = {OpenReview.net},
  address = {Singapore},
  numpages = {38},
  url = {https://openreview.net/forum?id=OJd3ayDDoF},
}

@inproceedings{kgent,
  author    = {Zheng, Yusheng and Yang, Yiwei and Chen, Maolin and Quinn, Andrew},
  title     = {{Kgent}: Kernel Extensions Large Language Model Agent},
  booktitle = {Proceedings of the ACM SIGCOMM 2024 Workshop on eBPF and Kernel Extensions},
  series    = {eBPF '24},
  year      = {2024},
  pages     = {30--36},
  publisher = {ACM},
  address   = {Sydney, NSW, Australia},
  doi       = {10.1145/3672197.3673434},
}

@misc{forkexplorecommit,
  title={Fork, Explore, Commit: {OS} Primitives for Agentic Exploration},
  author={Wang, Cong and Zheng, Yusheng},
  howpublished={arXiv preprint arXiv:2602.08199},
  year={2026},
  doi={10.48550/arXiv.2602.08199},
}

@misc{acrfence,
  title={{ACRFence}: Preventing Semantic Rollback Attacks in Agent Checkpoint-Restore},
  author={Zheng, Yusheng and Yang, Yiwei and Zhang, Wei and Quinn, Andi},
  howpublished={arXiv preprint arXiv:2603.20625},
  year={2026},
  doi={10.48550/arXiv.2603.20625},
}

@misc{agentcgroup,
  title={{AgentCgroup}: Understanding and Controlling {OS} Resources of {AI} Agents},
  author={Zheng, Yusheng and Fan, Jiakun and Fu, Quanzhi and Yang, Yiwei and Zhang, Wei and Quinn, Andi},
  howpublished={arXiv preprint arXiv:2602.09345},
  year={2026},
  doi={10.48550/arXiv.2602.09345},
}

@misc{agentry,
  author = {Wang, Cong},
  title = {{Agentry}: Orchestration and Memory for Multi-Agent Systems},
  howpublished = {\url{https://github.com/amtp-protocol/agentry}},
  year = {2025},
}

@misc{firejail,
  author = {{Firejail Project}},
  title = {{Firejail}: Linux Namespaces and Seccomp-bpf Sandbox},
  howpublished = {\url{https://github.com/netblue30/firejail}},
  year = {2026},
  note = {Accessed: 2026-05-23},
}

@misc{bubblewrap,
  author = {{Bubblewrap Project}},
  title = {{Bubblewrap}: Low-Level Unprivileged Sandboxing Tool},
  howpublished = {\url{https://github.com/containers/bubblewrap}},
  year = {2026},
  note = {Accessed: 2026-05-23},
}

@misc{camel,
  author = {Debenedetti, Edoardo and Shumailov, Ilia and Fan, Tianqi and Hayes, Jamie and Carlini, Nicholas and Fabian, Daniel and Kern, Christoph and Shi, Chongyang and Terzis, Andreas and Tram\`{e}r, Florian},
  title = {Defeating Prompt Injections by Design},
  howpublished = {arXiv preprint arXiv:2503.18813},
  year = {2025},
  doi = {10.48550/arXiv.2503.18813},
}

@misc{lethaltrifecta,
  author = {Willison, Simon},
  title = {The Lethal Trifecta for {AI} Agents: Private Data, Untrusted Content, and External Communication},
  howpublished = {\url{https://simonwillison.net/2025/Jun/16/the-lethal-trifecta/}},
  year = {2025},
}

@misc{dualllm,
  author = {Willison, Simon},
  title = {The Dual {LLM} Pattern for Building {AI} Assistants That Can Resist Prompt Injection},
  howpublished = {\url{https://simonwillison.net/2023/Apr/25/dual-llm-pattern/}},
  year = {2023},
}

@inproceedings{mbox,
  author    = {Kim, Taesoo and Zeldovich, Nickolai},
  title     = {Practical and Effective Sandboxing for Non-root Users},
  booktitle = {Proceedings of the 2013 USENIX Annual Technical Conference},
  series    = {USENIX ATC '13},
  year      = {2013},
  pages     = {139--144},
  publisher = {USENIX Association},
  address   = {San Jose, CA, USA},
  url       = {https://www.usenix.org/conference/atc13/technical-sessions/presentation/kim},
}

@inproceedings{garfinkel-traps,
  author    = {Garfinkel, Tal},
  title     = {Traps and Pitfalls: Practical Problems in System Call Interposition Based Security Tools},
  booktitle = {Proceedings of the Network and Distributed System Security Symposium (NDSS)},
  year      = {2003},
  numpages  = {15},
  publisher = {Internet Society},
  address   = {San Diego, CA, USA},
  url       = {https://www.ndss-symposium.org/ndss2003/traps-and-pitfalls-practical-problems-system-call-interposition-based-security-tools/},
}

@misc{nsjail,
  author       = {{\'S}wi{\k{e}}cki, Robert},
  title        = {{nsjail}: A Light-Weight Process Isolation Tool},
  howpublished = {\url{https://github.com/google/nsjail}},
  year         = {2024},
}

@misc{minijail,
  author       = {{The ChromiumOS Authors}},
  title        = {Minijail},
  howpublished = {\url{https://google.github.io/minijail/}},
  year         = {2024},
}

@inproceedings{bpfbox,
  author    = {Findlay, William and Somayaji, Anil and Barrera, David},
  title     = {{bpfbox}: Simple Precise Process Confinement with {eBPF}},
  booktitle = {Proceedings of the 2020 ACM SIGSAC Conference on Cloud Computing Security Workshop},
  series    = {CCSW '20},
  year      = {2020},
  pages     = {91--103},
  publisher = {ACM},
  address   = {Virtual Event, USA},
  doi       = {10.1145/3411495.3421358},
}

@misc{birdcage,
  author       = {{Phylum}},
  title        = {Birdcage: Cross-platform Embeddable Sandboxing},
  howpublished = {\url{https://github.com/phylum-dev/birdcage}},
  year         = {2023},
}

@misc{oci-runtime-spec,
  author = {{Open Container Initiative}},
  title = {{Open Container Initiative Runtime Specification}, Version 1.1.0},
  howpublished = {\url{https://oci-playground.github.io/specs-latest/specs/runtime/v1.1.0/oci-runtime-spec.html}},
  year = {2023},
}

\end{document}